\documentclass[fleqn,usenatbib]{mnras}

\usepackage[T1]{fontenc}

\DeclareRobustCommand{\VAN}[3]{#2}
\let\VANthebibliography\thebibliography
\def\thebibliography{\DeclareRobustCommand{\VAN}[3]{##3}\VANthebibliography}

\usepackage{graphicx}	% Including figure files
\usepackage{amssymb}	% Extra maths symbols
\usepackage[ruled,vlined]{algorithm2e}  % Algorithm formatting
\usepackage{fontawesome}    % For GitHub icon

\usepackage{newtxtext,newtxmath}
\title[Saliency mapping of galaxy morphologies]{Explaining deep learning of galaxy morphology with saliency mapping}

\author[P. Bhambra et al.]{
Prabh Bhambra,$^{1}$\thanks{E-mail: prabh.bhambra.12@ucl.ac.uk}
Benjamin Joachimi,$^{1}$
Ofer Lahav$^{1}$
\\
$^{1}$Department of Physics and Astronomy, University College London, Gower Street, London WC1E 6BT, UK
}

\date{Accepted XXX. Received YYY; in original form ZZZ}

\pubyear{2021}

\begin{document}
\label{firstpage}
\pagerange{\pageref{firstpage}--\pageref{lastpage}}
\maketitle

\begin{abstract}
We successfully demonstrate the use of explainable artificial intelligence (XAI) techniques on astronomical datasets in the context of measuring galactic bar lengths. The method consists of training convolutional neural networks on human classified data from Galaxy Zoo in order to predict general galaxy morphologies, and then using SmoothGrad (a saliency mapping technique) to extract the bar for measurement by a bespoke algorithm. We contrast this to another method of using a convolutional neural network to directly predict galaxy bar lengths. These methods achieved correlation coefficients of 0.76 and 0.59, and root mean squared errors of 1.69 and 2.10 respective to human measurements. We conclude that XAI methods outperform conventional deep learning in this case, which could be reasonably explained by the larger datasets available when training the models. We suggest that our XAI method can be used to extract other galactic features (such as the bulge-to-disk ratio) without needing to collect new datasets or train new models. We also suggest that these techniques can be used to refine deep learning models as well as identify and eliminate bias within training datasets.
\end{abstract}


%%%%%%%%%%%%%%%%%%%%%%%%%%%%%%%%%%%%%%%%%%%%%%%%%%

%%%%%%%%%%%%%%%%% BODY OF PAPER %%%%%%%%%%%%%%%%%%

\section{Introduction}
\label{sec:intro}

From autonomous vehicles to targeted advertising, machine learning (ML) algorithms have become more and more ubiquitous in society and a greater emphasis is being put on understanding the decisions they make. We call an algorithm interpretable if it can be understood by humans on its own without much analysis. Algorithms themselves vary in their interpretability, with the likes of decisions trees tending towards being more interpretable, while neural networks tend towards being some of the least interpretable \citep{expint_spectrum}. Despite their opacity, neural networks are widely used due to their relatively good performance compared to other ML algorithms. The field of explainable artificial intelligence (XAI) seeks to help explain the behaviour of difficult to interpret algorithms \citep{xai_review}.

ML has been widely adopted in the the field of astronomy, although to date relatively little thought has been given to the field of XAI. This paper aims to use XAI methods to generalise ML models trained on astronomical datasets. Due to astronomy's reliance on telescopic imaging, we considered datasets primarily consisting of images. Due to the richness of the data, the Galaxy Zoo project was chosen \citep{galaxy_zoo_2}. To our knowledge this is the first attempt at quantitatively using saliency mapping for image processing in astronomy, although there are recent cases of them being used qualitatively for spectral analysis \citep{saliency_spectra} and image processing \citep{saliency_21cm}. This paper intends to act as a proof of concept that these techniques are effective when applied to this field, and that they can be used to infer quantitative measurements.

Galaxy Zoo is a citizen science project in which volunteer participants are shown images of galaxies, and then asked a series of questions regarding their morphologies. Human classification of galaxy morphologies is extremely time consuming and requires expert level domain knowledge. However the use of citizen science platforms helps to reduce the reliance on these limited resources. Showing the same image to multiple participants can be used reach a crowd-sourced consensus of the galaxy morphology without requiring extreme time commitment from any single participant \citep{galaxy_zoo}.

There have been early applications of ML methods to aid in the classifications of galaxies \citep[e.g.][]{galaxies_humans_anns, ai_galaxy_class, min_span_tree}. Recently, the Galaxy Zoo project has provided a rich data source for the training of these ML models. Many implementations can be seen on the Kaggle competition ``Galaxy Zoo - The Galaxy Challenge'' in which participants were tasked to train an ML model on a subset of Galaxy Zoo data in order to predict galaxy morphologies. \citet{kaggle_winner} provided the winning submission to the contest by using convolutional neural networks (CNNs), a model architecture which is widely used in image processing due to their excellent performance \citep{cnns}. More recently, \citet{gz_to_des} used Galaxy Zoo data to train ML models which were then used to classify galaxies from the Dark Energy Survey. \citet{gz_bayesian_cnns} used Bayesian CNNs, which can more effectively learn from examples with uncertain labels, to predict posteriors for the morphologies of galaxies.

XAI methods can be used to produce saliency maps to show which areas in an image are most important to a CNN's classification of an image. This paper aims to use SmoothGrad, an XAI method developed to give visual explanations of CNNs, to produce saliency maps explaining which parts of an image are important in the image's classification as a barred galaxy. From these saliency maps the length of the bars can then be measured and compared to other methods of measurement. Galactic bars are of particular interest due to their ability to contribute to the redistribution of matter in a galaxy through transferring angular momentum between the disc, bulge, and any spiral arms or rings present \citep{bar_bulge, bar_ring}. Bar lengths are an especially good application of XAI due to the simplicity of their geometric structure. Galaxies are seen in 2D projection, and the highly linear structure of bars makes it straightforward to quantify the output of XAI methods and compare this quantification to other methods of measurement. This comparison can be used to demonstrate the predictive power of CNN classifications. We aim to show that XAI methods can be used to extract new information, which has never been presented to the CNN, from its classifications.

This paper is structured as follows: Section~\ref{sec:data} describes the data used throughout this paper, while Sections~\ref{sec:xai} and \ref{sec:model} describe the XAI methods used and the modelling process respectively. Section~\ref{sec:results} describes the results of comparing our methods to human measurements. Finally, Section~\ref{sec:conc} contains our conclusion.

\section{Data}
\label{sec:data}

The data used to train the models was taken from Kaggle's ``Galaxy Zoo - The Galaxy Challenge'' dataset\footnote{\url{https://www.kaggle.com/c/galaxy-zoo-the-galaxy-challenge/}}. This dataset consists of galaxy images each with dimensions $424 \times 424 \times 3$ (i.e. a height and width of 424 pixels, and 3 (red, green, and blue) colour channels). The images were partitioned into a training set and a test set. The training set consisted of 61,578 images with associated galaxy classification probability distributions. The probability distributions represent the proportion of Galaxy Zoo volunteers who associated the galaxy into each of the 37 classes. The test set consisted of 79,975 galaxy images without labels (as the testing of the models is handled solely by Kaggle's submission system). The galaxy images are composites of images taken by the Sloan Digital Sky Survey (SDSS) camera through its u, g, r, i, and z filters \citep{sdss}.

The Hoyle bar length catalogue \citep{bar_lengths} was collected as an extension to the Galaxy Zoo project. Galaxy Zoo participants were asked to measure the length of the bars of 3,150 barred galaxies by drawing a line on the images using a web interface. With multiple participants measuring the bars on each galaxy, participants were able to reproduce each other's measurements with a scatter of 17 per cent \citep{bar_lengths}. The bar length measurements in the Hoyle catalogue as well as the Galaxy Zoo images were scaled by the \texttt{petroR90\_r} value (the radius containing 90 per cent of the r-band Petrosian aperture flux) for the respective galaxy \citep{galaxy_zoo_2, bar_lengths}, hence this data was required to undo this scaling. These values were taken from the Galaxy Zoo metadata file hosted on the project's data page\footnote{\url{https://data.galaxyzoo.org/}}. The use of the bar length catalogue also required a full set of Galaxy Zoo 2 images \citep{gz_images}.

The Hoyle bar length catalogue also includes an error that introduces a proportional offset. The values in the catalogue are quoted in units of $h^{-1}~\rm{kpc}$, but are actually in units of $\rm{kpc}$ with $h$ assumed to be equal to $0.7$. Therefore, in order to use the catalogue in its intended units of $h^{-1}~\rm{kpc}$, we multiplied through by $h=0.7$ (Walmsley \& Geron, personal communication, 15 November 2021).

\section{Saliency mapping}
\label{sec:xai}

Saliency mapping is a term used to refer to a family of techniques seeking to explain the behaviour of CNNs. Saliency maps are heatmaps used to demonstrate the relative importance of parts of an image to the classification of that image into a certain class by the CNN in question \citep{smoothgrad}. These heatmaps are often computed by taking gradients from the output layer of the CNN to various other layers, in this way the resulting saliency map can be seen as visualising the sensitivity of an output to a given input \citep{saliency}.

Various methods were considered for use. Grad-CAM \citep{gradcam} and Grad-CAM++ \citep{gradcampp} take gradients from the output of the CNN to the penultimate convolution layer in order to assign importance to neurons within the CNN. This approach produces saliency maps which are able to locate features well but are unable to pull out finer detail within the image. Conversely, Guided Backpropagation \citep{gbp} manages to highlight features well, but is unable to strongly isolate important areas of the image. We ultimately decided to use SmoothGrad \citep{smoothgrad}, which provides explanations up to the pixel space of the input image. This allows SmoothGrad to pick out the finer detail in the image when explaining the behaviour of the CNN.

SmoothGrad builds on top of the idea of traditional saliency mapping \citep{saliency}, in which the importance ($L$) of a pixel's intensity ($x$) to the classification of class ($c$) is given by the gradient of the score of the relevant class ($y$) with respect to the pixel intensity:
\begin{equation}
    L^{c} (x) = \frac{\partial y^{c}}{\partial x}.
    \label{eq:saliency}
\end{equation}
$L^{c}$ is calculated for each pixel within the image and then a two-dimensional saliency map is constructed by stacking the individual values of $L^{c}$.

However, due to modern models combining many linear and non-linear (e.g. Rectified Linear Unit) operations, these gradients are locally unstable. Small fluctuations in the value of a pixel create visually similar images, however these can result in wildly different saliency maps \citep{smoothgrad}. SmoothGrad aims to address this issue by creating $N$ visually similar samples of the same image by adding Gaussian noise to each, and then averaging the saliency maps for each sample:
\begin{equation}
    \hat{L}^{c} (x) = \frac{1}{N} \sum^{N}_{i=1} L^{c} [x + \mathcal{N}(0, \sigma^{2})],
    \label{eq:smoothgrad}
\end{equation}
where $\mathcal{N}(0, \sigma^{2})$ is the probability density function for a Gaussian distribution with a mean of $0$ and a standard deviation of $\sigma^{2}$. Both $N$ and $\sigma^{2}$ are free parameters in the SmoothGrad algorithm.

\begin{figure*}
    \includegraphics[width=\textwidth]{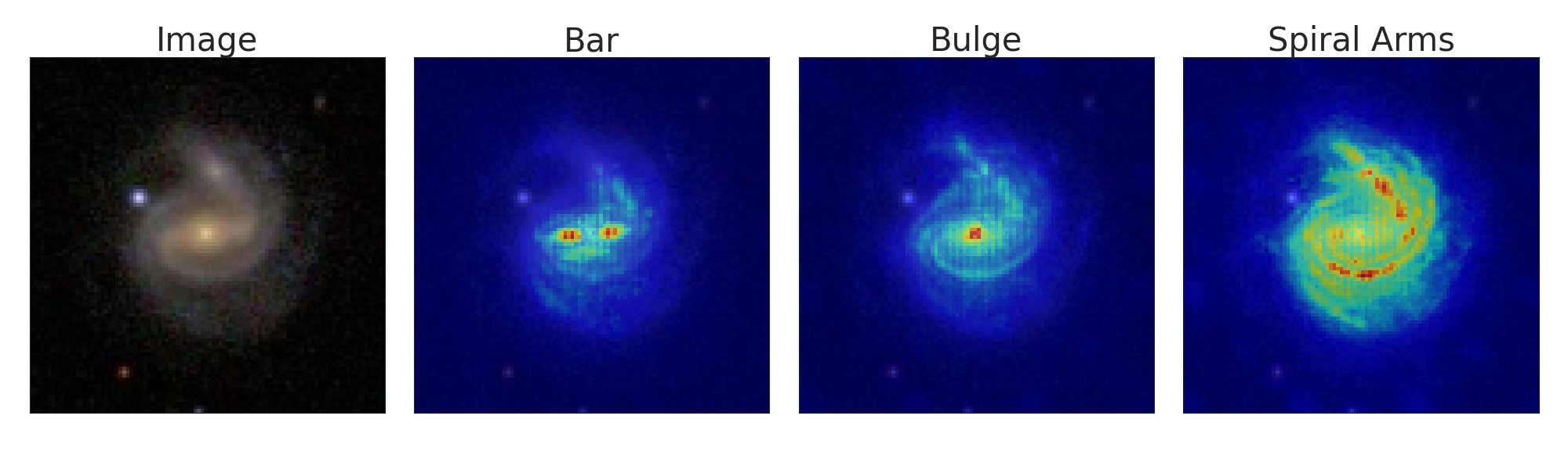}
    \caption{Example SmoothGrad saliency maps explaining the classification of a galaxy image (left) as a barred (centre-left) galaxy with a prominent bulge (centre-right) and spiral arms (right). Red pixels represent pixels with the highest importance $\hat{L}^{c}$, while blue represents those with the least importance. The saliency maps were generated using the models described in Section~\ref{subsec:model_architecture}. All images were cropped to the centre $100 \times 100$ pixels for this figure. The galaxy shown is identified by the SDSS Objid of 587722983363248237, which was chosen due to its clear demonstration of multiple features. Similar saliency maps for a random selection of galaxies can be found in Appendix~\ref{app:saliency_mapping}.}
    \label{fig:class_heatmaps}
\end{figure*}

Examples of SmoothGrad saliency maps can be found in Figure~\ref{fig:class_heatmaps}. The galaxy image was used to generate saliency maps explaining why models classified the galaxy as a barred galaxy with a prominent bulge and spiral arms. The saliency maps were compiled by averaging the individual saliency maps generated from each of the models described in Section~\ref{subsec:model_architecture}, using SmoothGrad parameters of $N=256$ and $S_{\rm{noise}}=0.1$ (where $S_{\rm{noise}}$ is a transformation of $\sigma^{2}$, see Section~\ref{subsec:model_hyperparameter}). This example demonstrates SmoothGrad's ability to pick out features of the galaxy when explaining the CNN's decisions to classify the image into various classes. It can be seen that SmoothGrad is successfully able to highlight the bar, bulge, and spiral arms in each successive saliency map.

\section{Machine learning model}
\label{sec:model}

\subsection{Data augmentation}
\label{subsec:model_augmentation}

Initial attempts at training models directly on unaugmented images resulted in poor convergence, therefore image augmentation was used. The images were down-sampled to dimensions of $224 \times 224 \times 3$ in order to reduce the impact of noise, and were also subjected to random augmentations during each epoch in order to reduce overfitting:

\begin{enumerate}
    \item random rotations in the range $0^{\circ} - 360^{\circ}$,
    \item random horizontal and vertical flips,
    \item random horizontal and vertical shifts in the range of 0-5 per cent of the image dimensions,
    \item random brightness scaling in the range of 70-130 per cent.
\end{enumerate}

\subsection{Model architecture}
\label{subsec:model_architecture}

In total, three different models were trained to predict galaxy morphologies, each based on a common image classification model (the VGG16 \citep{vgg16}, ResNet50v2 \citep{resnet50v2}, and Xception \citep{xception} architectures). Each model was created with a number of input nodes to match the size of the down-sampled images, and 37 (i.e. the number of Galaxy Zoo classes) nodes in the output layer. All models were implemented with the ``imagenet'' pre-trained weights from the Keras API for TensorFlow \citep{tensorflow}.

\subsection{Training}
\label{subsec:model_training}

The models were trained with the augmented images as the input and the Galaxy Zoo class probability distributions as the targeted output. 5 per cent of the training dataset was used for validation to monitor overfitting.

The models were trained in order to minimise the root mean squared error (RMSE) between the model's predictions and the Galaxy Zoo class probability distributions of galaxy morphologies. This was done by using an Adam optimiser \citep{adam}. The learning rate was also set to reduce by a factor of 0.5 if the loss on the validation dataset did not decrease over the course of five epochs. The models were set to be trained for up to 100 epochs; however training was halted in the case that the RMSE of the validation set did not decrease over the course of 10 epochs.

\begin{figure}
    \includegraphics[width=\columnwidth]{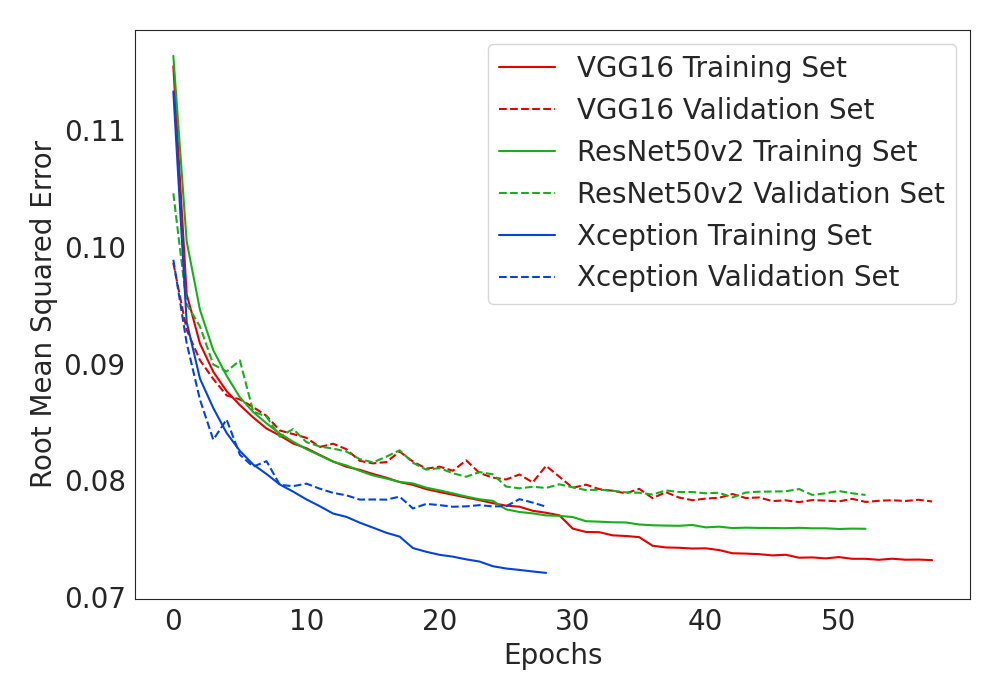}
    \caption{The learning curves for the three CNNs architectures trained. The losses for the VGG16 (red), ResNet50v2 (green), and Xception (blue) based models are shown for both the training (solid) and validation (dashed) sets.}
    \label{fig:learning_curves}
\end{figure}

The learning curves for the models can be found in Figure~\ref{fig:learning_curves}. The final metrics for each model can be found in Table~\ref{tab:model_evaluation} which shows the number of epochs each model trained for before stopping, as well as the RMSE on the training, validation, and test sets. The models were also ensembled by averaging the predictions of each individual model. The evaluation of the models on the test set was handled by Kaggle's submission system. Table~\ref{tab:model_evaluation} shows little variation in the performance of the models, which demonstrates the robustness of our training methods to model architecture choices. These models performed slightly worse than those trained by \citet{kaggle_winner} due to not using as intensive data augmentation practices.

\begin{table}
    \caption{RMSE on the training, validation, and test sets for each model architecture, including an ensembled model created by averaging the predictions of the other three models.}
    \label{tab:model_evaluation}
    \begin{tabular}{lcccc}
        \hline
        Model       & Epochs    & Train RMSE    & Val RMSE  & Test RMSE \\
        \hline
        VGG16       & 58        & 0.07315       & 0.07818   & 0.07939   \\
        ResNet50v2  & 53        & 0.07584       & 0.07874   & 0.08110   \\
        Xception    & 29        & 0.07206       & 0.07774   & 0.07933   \\
        Ensemble    & N/A       & N/A           & N/A       & 0.07747   \\
        \hline
    \end{tabular}
\end{table}

\subsection{Bar length measurement}
\label{subsec:model_bar_length_measurement}

We wanted to find a way to quantify how well SmoothGrad was able to explain the behaviour of the CNNs and how much information could be extracted from the CNN's classifications. Attempting to measure bar lengths from saliency maps was decided upon due to the availability of human measured bar lengths which could act as ``ground truth'' measurements. SmoothGrad was used to generate saliency maps highlighting the pixels within each galaxy in the Hoyle Bar Length catalogue that had the most importance (i.e. highest $\hat{L}^{c}$) in classifying the images as barred galaxies by the models. SmoothGrad has two hyperparameters \citep{smoothgrad}:

\begin{enumerate}
    \item $N$ - the number of samples to consider,
    \item $\sigma^{2}$ - the standard distribution of the Gaussian noise to be added to each sample.
\end{enumerate}

$N$ was set to 256. A greater number of samples would lead to less noisy saliency maps, but this would increase the required computing time significantly. The optimal value of $\sigma^{2}$ is dependent on the dataset being considered, therefore this was considered as a hyperparameter to be optimised.

In addition, elementwise multiplication of the saliency map with the input image is considered. This hyperparameter will be denoted as $X_{\rm{input}}$ and acts as a boolean flag. As we expect the bar of the galaxy to be brighter than the average brightness of the image, this should bias the saliency maps towards highlighting features of the galaxy rather than mistakenly highlighting any other artefacts or parts of the background.

Algorithm~\ref{algo:bar_length_measurement} was used to measure the bar length of a galaxy from its respective saliency map. The algorithm isolates the bar of the galaxy by zeroing out pixels in the saliency map below a specific threshold. This threshold is incremented until either the non-zero pixels are linearly distributed over the image with a Pearson correlation coefficient greater than a user-specified value, or the number of non-zero pixels is less than a user-specified required number of pixels. This algorithm requires two hyperparameters to be optimised:

\begin{enumerate}
    \item $T_{\rm{c}}$ the Pearson correlation threshold over which the non-zero pixels are considered linearly distributed;
    \item $T_{\rm{l}}$ - the lower bound on the number of non-zero pixels required.
\end{enumerate}

The bar lengths returned by Algorithm~\ref{algo:bar_length_measurement} are measured in units of pixels. However, Galaxy Zoo images are scaled \citep{galaxy_zoo_2} by $(\texttt{petroR90\_r} \times 0.02 \times \texttt{pixel\_scale})$ for each galaxy, where \texttt{pixel\_scale} is equal to $0.396~\rm{arcsec~pixel}^{-1}$ \citep{sdss}. Therefore our measured bar lengths were later multiplied by this value to convert the measurements to units of $h^{-1}~\rm{kpc}$. The measured lengths also had to be multiplied by a factor of $(424/224)$ in order to undo the image down-sampling described in Section~\ref{subsec:model_augmentation}.

\begin{algorithm}
    \SetAlgoLined
    \DontPrintSemicolon
    \SetKw{KwOr}{or}
    \SetKw{KwBy}{by}
    \SetKwComment{Comment}{}{}
    \KwIn{saliencyMap, $T_{\rm{c}}$, $T_{\rm{l}}$}
    \KwOut{barLength}
    \For{$T_{h}$ = 0.5 \KwTo 1.0 \KwBy 0.0001}{
        x, y = where(saliencyMap < $T_{h}$)\;
        corr = corr(x, y)\;
        size = len(x)\;
        \If{corr > $T_{\rm{c}}$ \KwOr size < $T_{\rm{l}}$}{
            break\;
            }
        }
    x, y = where(saliencyMap > $T_{h}$)\;
    barLength = 0\;
    \For{i = 0 \KwTo len(x)}{
        \For{j = 0 \KwTo len(x)}{
            length = $\sqrt{(x[i] - x[j])^{2} + (y[i] - y[j])^{2}}$\;
            \If{length > barLength}{
                barLength = length\;
            }
        }
    }
    \Return barLength\;
    \caption{Measuring the bar length from a SmoothGrad saliency map. \protect\footnotemark}
    \label{algo:bar_length_measurement}
\end{algorithm}

\footnotetext{The corr() function calculates the Pearson correlation coefficient between two variables, and the where() function returns the indices of an array at which the condition is evaluated as True.}

\begin{figure*}
    \includegraphics[width=0.75\textwidth]{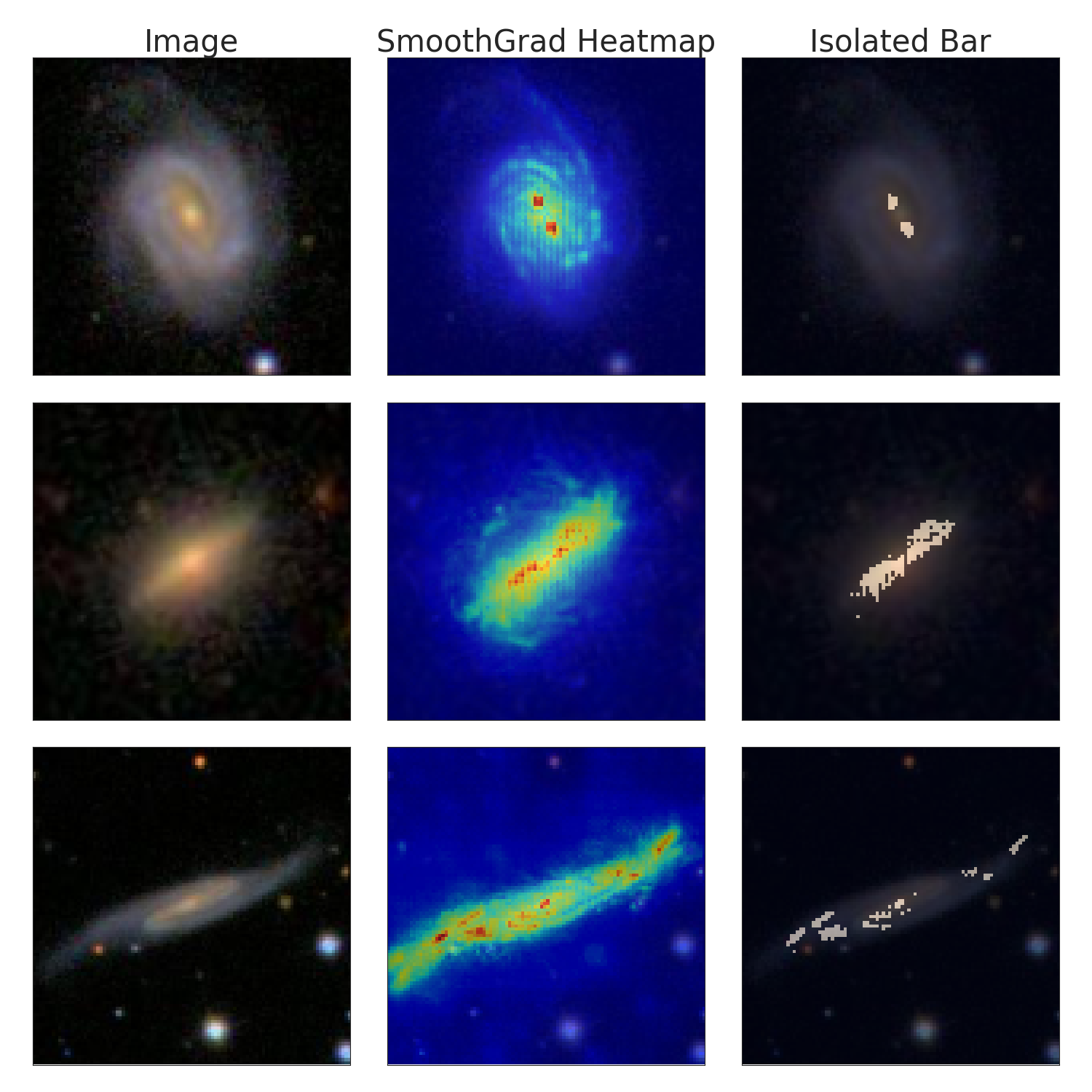}
    \caption{Examples of isolating the bar from galaxy images for three different galaxies. The leftmost column shows the initial image cropped to the centre $100 \times 100$ pixels. The middle column shows a saliency map for each galaxy highlighting the pixels most important in classifying the images as barred galaxies. The rightmost column shows the use of Algorithm~\ref{algo:bar_length_measurement} in isolating the bar from the relevant saliency maps. The first two examples were chosen at random, however the bottom example was chosen to present a case where this method fails (discussed in Section~\ref{subsec:results_bar_length_measurements}). The galaxies shown are identified by their SDSS Objids 588848901007933476, 588017977830342894, and 588016878291124448 from top to bottom. Similar saliency maps for a random selection of galaxies can be found in Appendix~\ref{app:bar_lengths}.}
    \label{fig:barlength_heatmaps}
\end{figure*}

\subsection{Hyperparameter optimisation}
\label{subsec:model_hyperparameter}

Overall, optimising the method of measuring bar lengths requires searching the space of four hyperparameters: $X_{\rm{input}}$, $\sigma^{2}$, $T_{\rm{c}}$, and $T_{\rm{l}}$. The values of hyperparameters were chosen as follows:

\begin{enumerate}
    \item $X_{\rm{input}} \in \{\rm{True}, \rm{False}\}$;
    \item $S_{\rm{noise}} = \sigma / (x_{\rm{max}} - x_{\rm{min}}) \in \{0.1, 0.15, 0.2, 0.25\}$;
    \item $T_{\rm{c}} \in \{0.7, 0.75, 0.8, 0.85, 0.9\}$;
    \item $T_{\rm{l}} \in \{10, 20, 30, 40, 50\}$,
\end{enumerate}
where $S_{\rm{noise}}$ is a transformation of $\sigma^{2}$, and $x_{\rm{max}}$ and $x_{\rm{min}}$ are the values of the most and least intense pixels in the image being evaluated.

The Hoyle bar length catalogue was separated into two sets, a validation set of 1,050 galaxies used for hyperparameter selection, and a test set of 2,100 galaxies used for final evaluation. The measured bar lengths were then compared to the \texttt{length\_scaled} values from the Hoyle bar length catalogue. The catalogue values had been scaled by the \texttt{petroR90\_r} value for each galaxy \citep{bar_lengths}, therefore this scaling was undone to ensure the values being used were in units of $h^{-1}~\rm{kpc}$.

Bar lengths were measured for each combination of hyperparameters using the following methods:

\begin{enumerate}
    \item saliency maps generated from the VGG16 based model;
    \item saliency maps generated from the ResNet50v2 based model;
    \item saliency maps generated from the Xception based model;
    \item saliency maps generated by elementwise averaging of the saliency maps from (i), (ii), and (iii);
    \item averaging the bar lengths found by methods (i), (ii), and (iii).
\end{enumerate}

Table~\ref{tab:best_correlations_val} shows the Pearson correlation coefficients ($\rho$) between the measured bar lengths and the catalogue values for the optimal set of hyperparameters of each method. The hyperparameters were optimised in order to maximise $\rho$, which is used as a measure of predictiveness.

\begin{table}
    \caption{Best Pearson correlation coefficients ($\rho$) found for the validation set, and required hyperparameters for each method (for all, the hyperparameters $X_{\rm{input}}$ and $S_{\rm{noise}}$ were found to be False and 0.1 respectively).}
    \label{tab:best_correlations_val}
    \begin{tabular}{lccc}
        \hline
        Method                  & $T_{\rm{c}}$  & $T_{\rm{l}}$  & $\rho$    \\
        \hline
        VGG16                   & 0.9           & 40            & 0.68      \\
        ResNet50v2              & 0.85          & 30            & 0.71      \\
        Xception                & 0.8           & 40            & 0.67      \\
        Average Saliency Maps   & 0.85          & 30            & 0.75      \\
        Average Measurements    & 0.85          & 40            & 0.76      \\
        \hline
    \end{tabular}
\end{table}

\subsection{Control model}
\label{subsec:model_control}

We also trained a CNN in order to directly predict the bar lengths of galaxies in the Hoyle catalogue from their respective galaxy images. Section~\ref{subsec:model_training} showed the the Xception architecture performed best at classifying galaxies, hence this architecture was chosen again to directly predict bar lengths. The same image augmentation methods were used as in Section~\ref{subsec:model_augmentation}, however the number of output nodes in the model was changed to one. The model was trained to minimise the RMSE of its predictions with respect to the \texttt{length\_scaled} value from the catalogue. The model was trained for 100 epochs with early stopping implemented such that training was ceased if there was no improvement over the course of 10 epochs.

For the purpose of training this model, the 3,150 galaxies in the catalogue were randomly split into training, validation, and test sets containing 64, 16, and 20 per cent of the images, respectively. Training was halted after 36 epochs at which the RMSE on the validation set was 0.10203.

\section{Results}
\label{sec:results}

\subsection{Bar length measurements}
\label{subsec:results_bar_length_measurements}

Table~\ref{tab:best_correlations_val} shows that the highest $\rho$ on the validation set were gained from the ``Average Measurements'' method. Therefore, this method was chosen to be evaluated on the test set of the catalogue (and will hereafter referred to as the ``Explainable Deep Learning'' method). Evaluation of this method on the test set was performed twice to reduce and quantify any uncertainty due to the Gaussian noise added as part of the SmoothGrad algorithm. The measured bar lengths are computed as the mean of the bar lengths found during each run. Performance was shown to be similar between the validation and test sets, with measured bar lengths having $\rho = 0.76$, and an RMSE of 1.69 respective to human measured values (using hyperparameters: $X_{\rm{input}} = \rm{False}$, $S_{\rm{noise}} = 0.1$, $T_{\rm{c}} = 0.85$, and $T_{\rm{l}} = 40$).

In comparison, the CNN described in Section~\ref{subsec:model_control} (hereafter referred to as the ``Direct Deep Learning'' method) only managed to achieve $\rho = 0.59$, and an RMSE of 2.10 between the measured and catalogue bar length values on the model's test set. This relatively poor performance could reasonably be explained by the much smaller size of the bar length catalogue used to train the model, in comparison to the size of the full Galaxy Zoo data used to train the other models. Although we could decouple the size of the dataset from the method, we feel like this would be intentionally handicapping XAI methods as one of their distinct advantages is that they do not require specialised data which is often harder to collect.

In most cases XAI methods are able to correctly highlight the bar of a galaxy image and measure it accordingly, as can be seen in the middle row of Figure~\ref{fig:barlength_heatmaps}. XAI methods can result in a few extreme outliers as seen in the bottom row of Figure~\ref{fig:barlength_heatmaps}. In this case the CNN mistakenly sees the entire galaxy as a bar due to the galaxy's inclination in the image, resulting in a vastly overestimated bar length measurement. The ``double blob'' structure in the saliency maps for the first example in Figure~\ref{fig:barlength_heatmaps} can be explained by the model not requiring the use of the prominent bulge in classifying these galaxies as barred, the bulge is not unique to barred galaxies and hence holds little importance in the model's classification.

We did not expect machine learning methods to be able to reproduce absolute values of human bar length measurements. Defining the start and end points of the bar is a difficult task for both humans and machines, so we expect differences between the two methods. However, a strong correlation between the two methods would be considered a success. Linear regression fits were performed by Scikit-learn \citep{scikit_learn} between the Hoyle bar length catalogue and the measured lengths for both the ``Explainable Deep Learning'' and ``Direct Deep Learning'' methods. The regression fits were forced to go through the origin, and were weighted by the uncertainty associated with each data point. For both fits the error on the Hoyle bar lengths were set to 17 per cent as specified in \citet{bar_lengths}. Errors for the bar lengths measured by the ``Explainable Deep Learning'' method were estimated using the spread of the measured bar lengths from each run over the test set. The spread for each galaxy ($\delta^{g}$) was computed as $\delta^{g} = \lvert l_{1} - l_{2} \rvert$, where $l_{1}$ and $l_{2}$ are the first and second measurements taken. The galaxies were then binned into deciles by their measured bar lengths. $\delta^{g}$ was averaged over all galaxies in each decile, and this average was used as the error for all the galaxies in the decile. Errors for the predictions of the ``Direct Deep Learning'' method were not quantified.

\begin{figure}
    \includegraphics[width=\columnwidth]{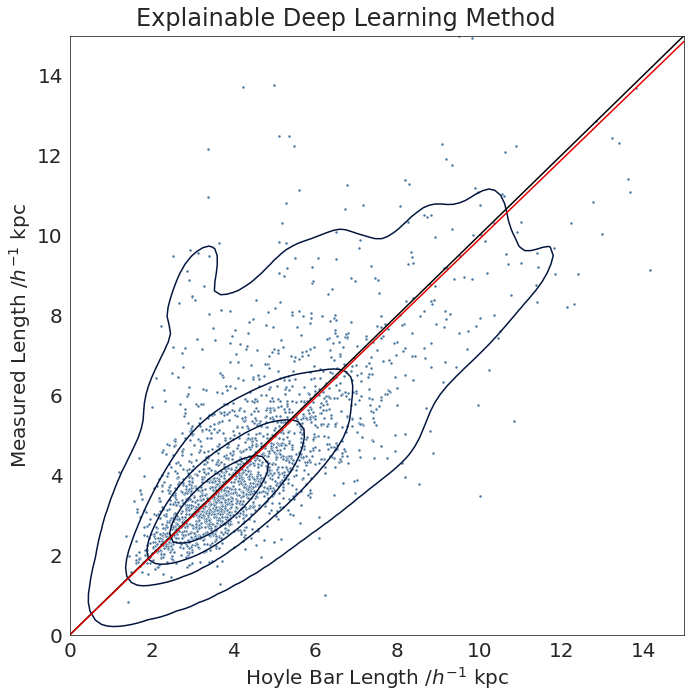}
    \includegraphics[width=\columnwidth]{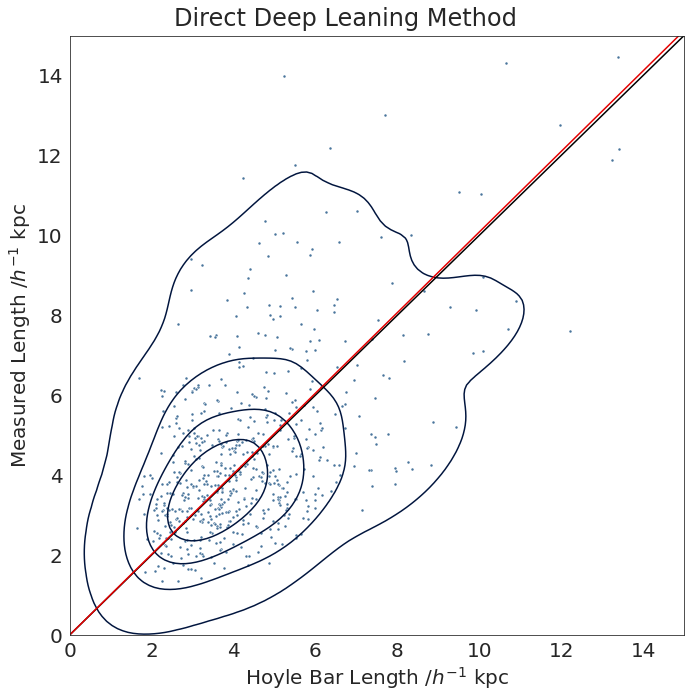}
    \caption{Bar lengths measured by the ``Explainable Deep Learning'' (top) and ``Direct Deep Learning'' (bottom) methods vs the Hoyle bar length catalogue for each method's test dataset. Weighted regression fits are also shown (red lines), as well as the one-to-one relation (black lines). A kernel density estimate overlay (dark blue contours) is provided to help illustrate the distribution of data points in each panel and error bars are not displayed for the sake of clarity. The regression fits for both are close to unity, however the ``Explainable Deep Learning'' has a smaller scatter than that of the ``Direct Deep Learning'' method indicating stronger predictive power.}
    \label{fig:barlength_scatter}
\end{figure}

Figure~\ref{fig:barlength_scatter} shows that the ``Explainable Deep Learning'' method not only correlates strongly with the catalogue values but also matches the values extremely well, as demonstrated by the close relation between the regression fit and the one-to-one relation. Although the regression fit for the ``Direct Deep Learning'' method is also close to unity, it is important to remember that the most important metric for evaluating the performance of these methods is $\rho$. The distribution of the data points on the ``Direct Deep Learning'' plot has more scatter than for ``Explainable Deep Learning'' plot, and is thus it is a less predictive method.

\subsection{Misclassification}
\label{subsec:results_misclassification}

\begin{figure}
    \includegraphics[width=\columnwidth]{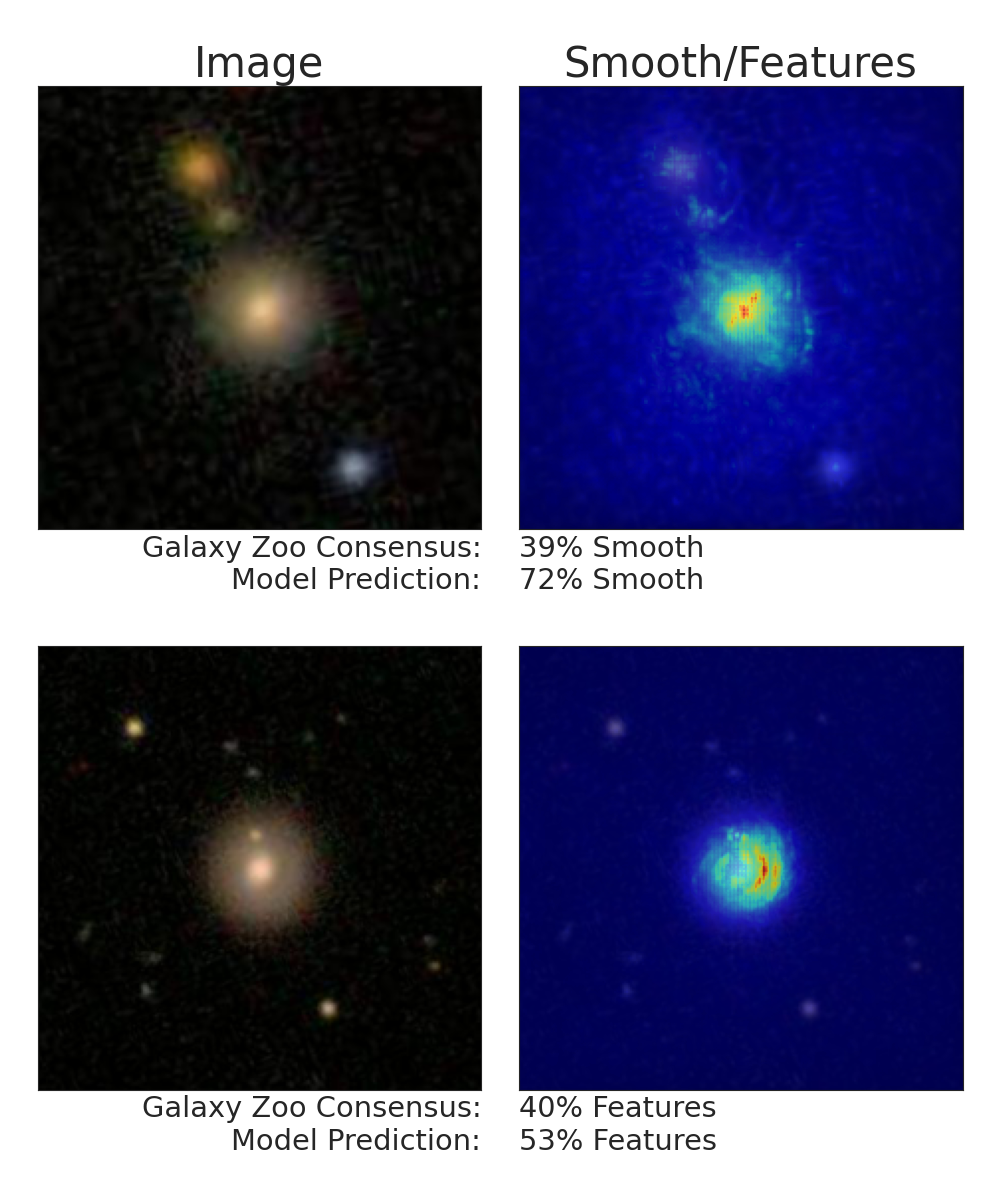}
    \caption{Two examples where the machine learning prediction disagrees with the Galaxy Zoo citizen science consensus. The image (left column) is shown along with SmoothGrad saliency maps explaining the model's classification of this image (right column). The top row shows a galaxy where the Galaxy Zoo consensus is that of a galaxy with features and structure, while the bottom row shows an example where the consensus is that of a smooth galaxy. Our model disagrees with both of these. The Galaxy Zoo consensus as well as the model predictions probabilities are given for both examples. The galaxies shown are identified by the filenames 102271.jpg and 120250.jpg from the Kaggle dataset.}
    \label{fig:misclassification_heatmaps}
\end{figure}

Throughout this study, we came across many instances where the XAI model predictions differed from the Galaxy Zoo consensus (i.e. the model's predicted answer to a Galaxy Zoo question was different to the answer with the highest share of votes amongst Galaxy Zoo participants). For the entire Kaggle dataset we observed that the XAI model predictions on whether a galaxy has any discernible features differed from the Galaxy Zoo consensus 8 per cent of the time. We present two examples in Figure~\ref{fig:misclassification_heatmaps} in which our ensembled models predict different morphologies to the consensus. These examples were chosen in order to present cases both where the model predicted features where Galaxy Zoo participants did not, and vice versa.

The top row shows a galaxy which our XAI model classifies as smooth (i.e. no bar, spiral arms, or other structure present) with 72 per cent confidence. Contrarily, only 39 per cent of Galaxy Zoo participants classified it as a smooth galaxy, with most classifying it as a having features (and some even classifying it as a galaxy with spiral arms). We can only speculate that these classifications are due to the neighbouring objects in the image (although Galaxy Zoo participants are instructed to only focus on the object in the centre of the image). The related saliency map shows the models being unable to pick out structure in this galaxy apart from a slight overdensity in the centre of the image. The bottom row shows a galaxy where 40 per cent of Galaxy Zoo participants have classified it as a smooth galaxy. Again, our model disagrees with this with a confidence of 53 per cent, and we can see from the saliency map that the model confidently picks out what looks like either a ring structure or dust lane in the galaxy image.

Looking at these examples, we agree more with the XAI model predictions than the Galaxy Zoo consensus. The SmoothGrad saliency maps shown in Figure~\ref{fig:misclassification_heatmaps} can help us understand why the models make their decisions, however we can only speculate as to why the Galaxy Zoo participants classified these examples in the way they did. It is possible to ask participants for their reasoning via the Galaxy Zoo Forum\footnote{\url{https://www.galaxyzooforum.org/}, although this would be extremely time consuming to do for a large number of galaxies and does not guarantee responses from the participants who provided the responses in the dataset.} This brings up an interesting scenario in which the citizen science method of classifying galaxies is less easily explained than deep learning methods.

\section{Conclusions}
\label{sec:conc}

Although machine learning (ML) is widely used in astronomy, we believe that we are the first to quantitatively demonstrate the effectiveness of saliency mapping within the field. We used SmoothGrad to extract bar length measurements from convolutional neural networks (CNN) trained on Galaxy Zoo data. This method of measuring bar lengths correlated with human measurements with a Pearson correlation coefficient of 0.76 and achieved a root mean squared error of 1.69. This vastly outperformed a CNN trained specifically to predict bar lengths, which was only able to achieve a correlation of 0.59 and root mean squared error of 2.10, with respect to human measurements.

The superiority of explainable artificial intelligence (XAI) methods could reasonably be explained due to the larger dataset that was used to train the ML models. The Galaxy Zoo dataset used contained 61,578 galaxy images for the training of models which can predict galaxy morphologies. In comparison, only 2,520 galaxy images were used in the training of the model tasked with directly predicting galactic bar lengths. The larger size of the Galaxy Zoo dataset allowed more general learning of galactic features and helped increase the predictiveness of the XAI based methods. A smaller dataset would reduce the performance of the initial model \citep{ml_benchmarking} and thus would reduce the ability of XAI methods to extract galactic features.

The major advantage of utilising XAI methods is that information which the model was never presented with can be extracted from the model predictions. This greatly reduces the reliance of collecting data in order to learn features of galaxies. We focused on bar lengths, however we could also extract other galactic features (such as the bulge-to-disk ratio) from the same models. Traditional deep learning methods would require a new and extensive dataset in order to be able to predict these features.

We also briefly discussed the difficulties in reconciling differences between the ML model predictions and the Galaxy Zoo consensus. One possible path for future work is to use models trained only on galaxies where there is an extremely strong consensus between Galaxy Zoo participants \citep[see e.g.][]{gz_agreement_model}, models designed to handle galaxies with uncertain labels \citep[see e.g.][]{gz_bayesian_cnns}, or to use pretrained models \citep[see e.g.][]{zoobot}. This could improve the ML model predictions, and lead to cleaner saliency maps and more accurate bar length measurements. Another possibility would be to use the saliency maps in order to identify and eliminate bias from the training dataset, which would lead to a higher quality dataset for training models.

We conclude that SmoothGrad was able to explain the decisions made by the models in a way that could be quantified and compared favourably to human action. This demonstrates that saliency mapping was successful in explaining the ML models used in this paper within the context of bar lengths, and could see a wide range of applications within the field of astronomy.

\section*{Acknowledgements}

The authors would like to thank A. Bevan for drawing our attention to saliency mapping techniques, and M. Walmsley for their insightful and helpful feedback. PB is supported by the STFC UCL Centre for Doctoral Training in Data Intensive Science. BJ acknowledges support by an STFC Consolidated Grant, grant no. ST/V000780/1. OL acknowledges STFC Consolidated Grant ST/R000476/1 and a Visiting Fellowship at All Souls College, Oxford.

\section*{Data Availability}

All code required to reproduce this work will be made publicly available on acceptance. The code includes a readme detailing the steps required to reproduce the study, including downloading all data and setting the seeds used when randomly splitting the datasets and augmenting training data. Please note that the version of the saliency mapping package used does not support the use of random seeds, however the high number of samples used in the SmoothGrad algorithm should minimise the impact of this. \href{https://github.com/prabhbhambra13/xai_bar_lengths/}{GitHub \faGithub}.

%%%%%%%%%%%%%%%%%%%%%%%%%%%%%%%%%%%%%%%%%%%%%%%%%%



\bibliographystyle{mnras}
\bibliography{bibliography}

%%%%%%%%%%%%%%%%%%%%%%%%%%%%%%%%%%%%%%%%%%%%%%%%%%

%%%%%%%%%%%%%%%%% APPENDICES %%%%%%%%%%%%%%%%%%%%%

\appendix
\section{Randomised Saliency Mapping Examples}
\label{app:saliency_mapping}

Here we present further examples similar to Figure~\ref{fig:class_heatmaps}, of SmoothGrad saliency maps for three chosen classes on five randomly selected galaxies. The galaxies selected are from the Kaggle dataset used to train the CNNs in Section~\ref{subsec:model_architecture}, with the condition that the selected galaxy must have at least a 50 per cent of voters classifying it as having a bar, a bulge, and spiral arms.

\begin{figure*}
    \includegraphics[width=\textwidth]{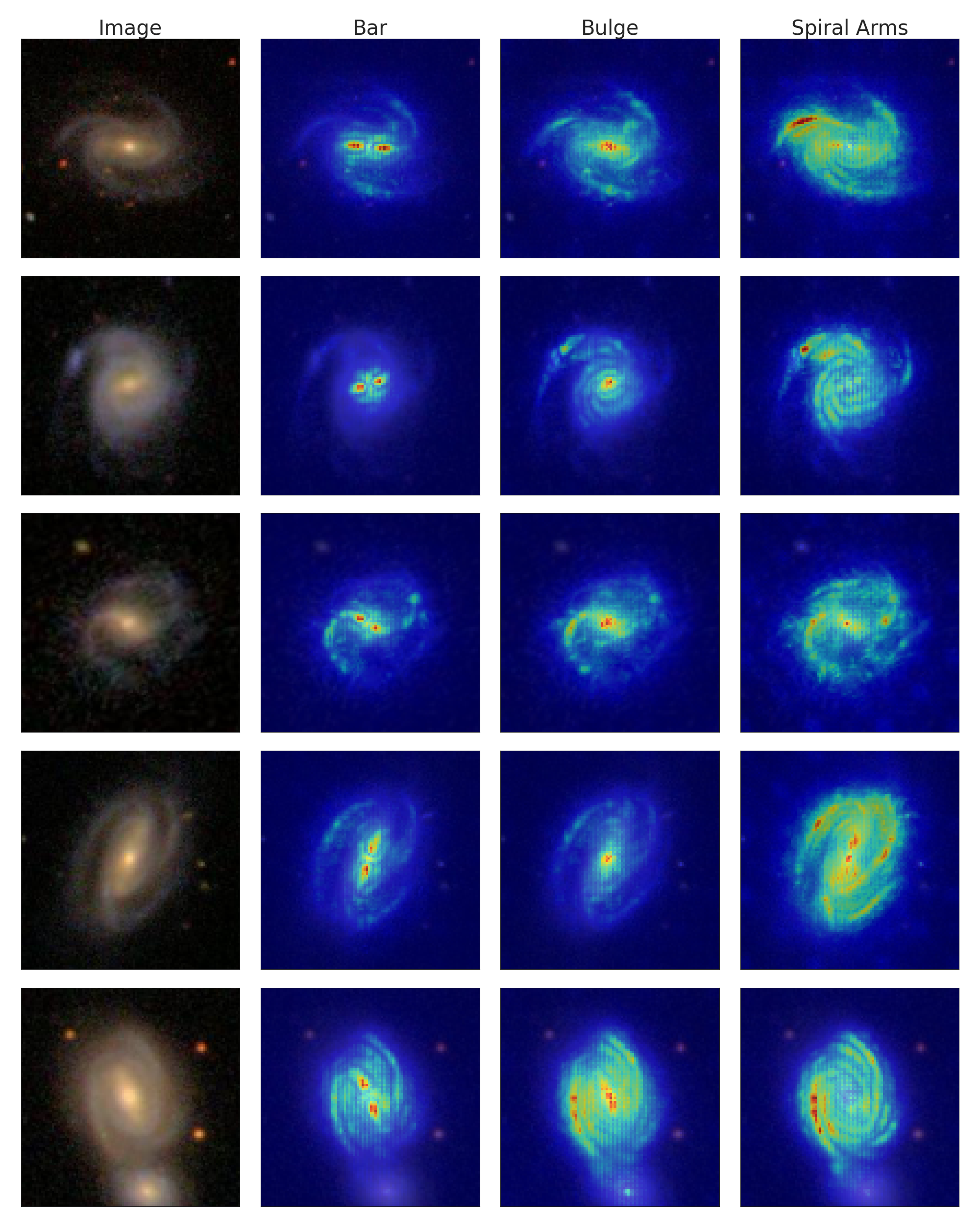}
    \caption{Similar to Figure~\ref{fig:class_heatmaps}, but with examples randomly selected from the Kaggle dataset, with the conditions that the selections must have at least 50 per cent of voters classifying it as having a bar, a bulge, and spiral arms.}
    \label{fig:appendix_a}
\end{figure*}

\section{Randomised Bar Length Measurement Examples}
\label{app:bar_lengths}

Here we present further examples similar to Figure~\ref{fig:barlength_heatmaps}, of SmoothGrad saliency maps isolating the bar of five randomly selected galaxies. The galaxies selected are from the test set of the Hoyle catalogue used to evaluate the performance of the XAI based methods in Section~\ref{subsec:results_bar_length_measurements}.

\begin{figure*}
    \includegraphics[width=0.75\textwidth]{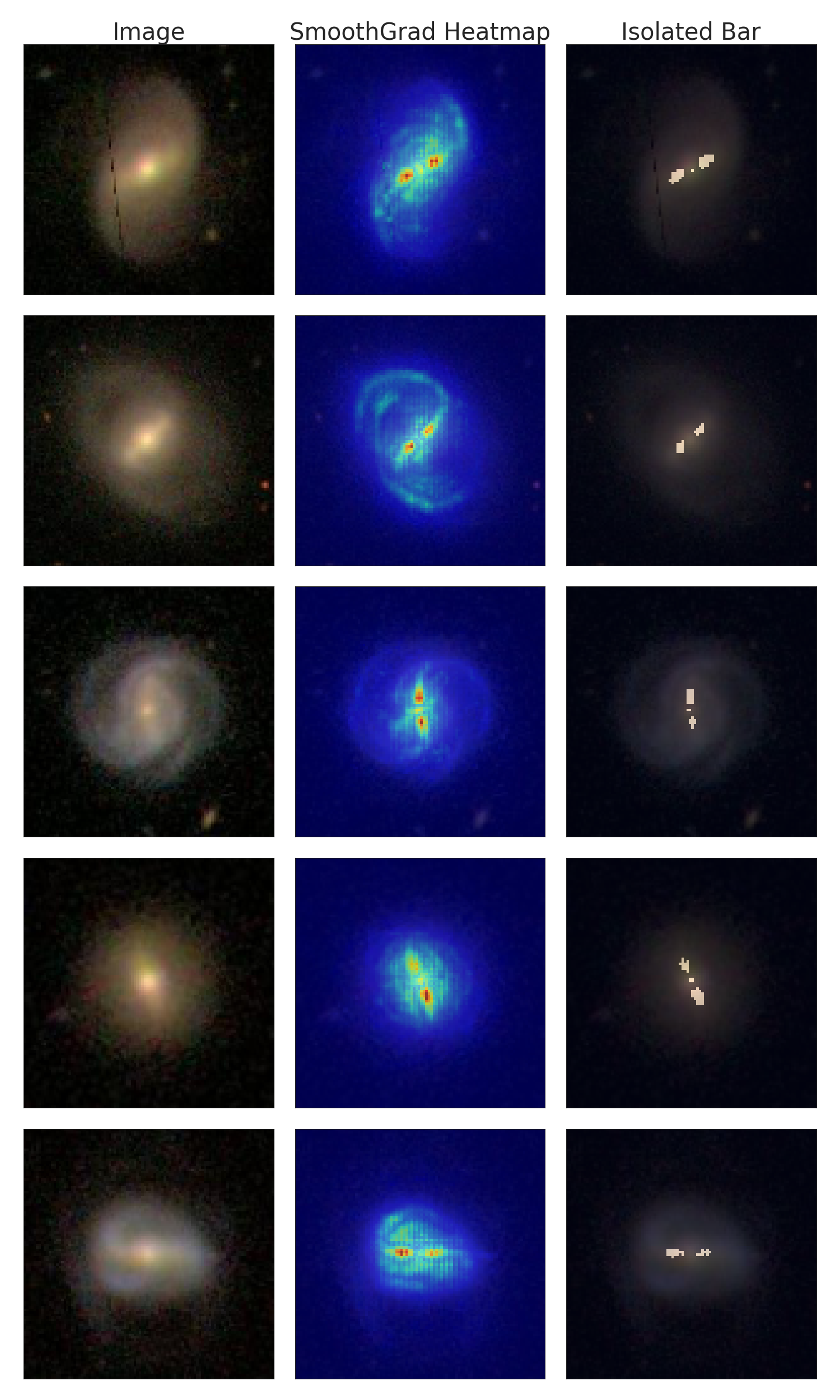}
    \caption{Similar to Figure~\ref{fig:barlength_heatmaps}, but with examples randomly selected from the Hoyle catalogue.}
    \label{fig:appendix_b}
\end{figure*}

%%%%%%%%%%%%%%%%%%%%%%%%%%%%%%%%%%%%%%%%%%%%%%%%%%

% Don't change these lines
\bsp	% typesetting comment
\label{lastpage}
\end{document}